\documentclass[prl,twocolumn,amsmath,amssymb]{revtex4}

\usepackage{color,epsfig,rotating,graphicx,subfigure}
\usepackage{amsmath,amssymb,amscd,afterpage}

\usepackage[latin1]{inputenc}
\usepackage[english]{babel}

\usepackage{dsfont}
\usepackage{textcomp}

\renewcommand{\Re}{{\rm Re}}
\renewcommand{\Im}{{\rm Im}}

\newcommand{\ri}{{\rm i}}
\newcommand{\re}{{\rm e}}
\newcommand{\rd}{{\rm d}}

\newcommand{\rs}{{\rm s}}
\newcommand{\rp}{{\rm p}}

\begin{document}

\title{Giant Spin Hall Effect in Single Photon Plasmonics}

\author{G.\ S.\ Agarwal}

\affiliation{Department of Physics, Oklahoma State University, Stillwater, Oklahoma 74078, USA}

\author{S.-A. Biehs}

\affiliation{Institut f\"{u}r Physik, Carl von Ossietzky Universit\"{a}t,
D-26111 Oldenburg, Germany.}

\date{\today}

           
\begin{abstract}
We show the existence of a very large spin Hall effect of light (SHEL) in single photon 
plasmonics based on spontaneous emission and the dipole-dipole interaction initiated energy 
transfer (FRET) on plasmonic platforms. The spin orbit coupling inherent in Maxwell equations 
is seen in the conversion of $\sigma^+$ photon to $\sigma^-$ photon. The FRET is mediated by the 
resonant surface plasmons and hence we find very large SHEL. We present explicit results 
for SHEL on both graphene and metal films. We also study how the splitting of the surface 
plasmon on a metal film affects the SHEL. In contrast to most other works which deal with 
SHEL as correction to the paraxial results, we consider SHEL in the near field of dipoles 
which are far from paraxial.
\end{abstract}

\maketitle
In recent years the SHEL has attracted considerable 
attention~\cite{OnadaEtAl2004,KavokinEtAl2005,Li2009,Bliokh}. Several experiments 
have confirmed the existence of the 
effect~\cite{CourtialEtAl1998,HostenKwiat2008,HaefnerEtAl2009,BonnetEtAl2001,HermosaEtAl2011,GorodetskiEtAl2008}. 
The simplest version of the effect is known as Federov-Imbert effect~\cite{Federov1955,Imbert1972,Berry2011} and 
is seen prominently as a polarization dependent transverse shift on the reflected beam at a 
dielectric interface. The shift occurs relative to the prediction of geometrical optics. The light 
beam is known to carry both orbital and spin angular momentum~\cite{AllenBarnettPadgett,YaoEtAl2011}
and the spin-orbit coupling is inherent in the vectorial Maxwell equations. When light travels across 
the interface between the two dielectrics then the orbital angular momentum changes which then implies 
change in the polarization so that the total angular momentum is conserved~\cite{SchwartzDogariu2006}. 
The magnitude of the shift is generally quite small. However, for reflection from metal surfaces shifts of 
the order of wavelength have been reported~\cite{HermosaEtAl2011}. The metallic gratings even yield 
larger result~\cite{BonnetEtAl2001} due to excitation of surface plasmons. The SHEL appears in a variety of 
other ways~\cite{KavokinEtAl2005,CourtialEtAl1998,HaefnerEtAl2009}.

Motivated by the recent progress in the observation of very large enhancement in spontaneous 
emission~\cite{GSA1975,RussellEtAl2012,OkamatoEtAl2008,CangEtAl2011,Eghlidi2009,LeeEtAl2011,NealEtAl2011,RousseauEtAl2009}, 
we propose the existence of giant SHEL in single photon plasmonics. Especially 
fabricated plasmonic structures (PS)~\cite{GuEtAl2010,GuEtAl2012} have been shown to be especially 
suitable for enhancement of fluorescence. We consider the F\"{o}rster energy transfer~\cite{AndrewBarnes2004,Blum2012} between 
two atoms located on a plasmonic surface. The energy transfer is mediated by plasmons. The SHEL 
arises as we demonstrate as a clear conversion of a, say, $\sigma^+$ photon into a $\sigma^-$ photon which is 
absorbed by the second atom leading to preparation of the second atom in a state which is orthogonally 
polarized to the state of the atom which produced the photon in the first place. The conversion is 
accompanied by the generation of the two units of orbital angular momentum. The giant SHEL can be detected 
by monitoring the population of the orthogonally polarized excited state of the second atom. We trace the 
existence of the effect to the fact that the dipole field is far from a paraxial field and in fact contains 
all the Fourier components including the evanescent waves. It has been shown that the image of a 
dipole in far field is displaced~\cite{ArnoldusEtAl2008,Berry2011}. Further Maxwell equations imply that the vectorial 
properties of the fields change on propagation. We demonstrate giant SHEL with explicit results 
for FRET on graphene and metallic platforms. The atoms on nano fibers~\cite{KienEtAl2005,GobanEtAl2012} would 
be another prominent system which would lead to large SHEL. Our proposal constitutes a new generation 
of possible experiments for large SHEL. 

\begin{figure}[Hhbt]
  \epsfig{file = 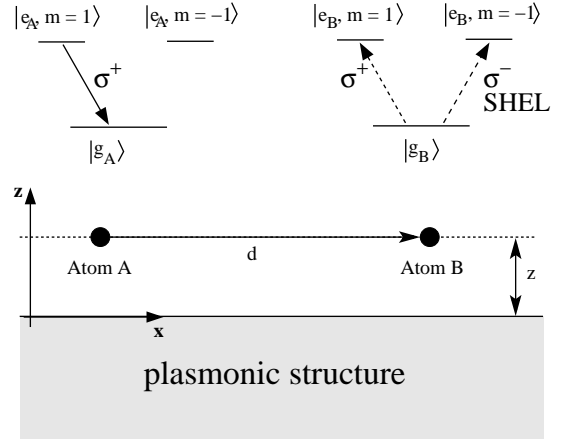, width = 0.4\textwidth}
  \caption{\label{Fig:Atoms} Sketch of the excitation transfer between atom $A$ and $B$.}
\end{figure}

Let us consider the radiative interaction and the energy transfer between two three-level atoms
$A$ and $B$ as depicted in Fig.~\ref{Fig:Atoms}. The transfer of the excitation of
atom $A$, say, to atom $B$ occurs via spontaneous emission from atom $A$ and
absorption of the emitted photon by atom $B$. Thus the single photon excitation 
transfer is due to the interaction with the vacuum field $\mathbf{E}$ in presence of
the plasmonic environment. For concreteness we assume that atom $A$ has been prepared
by pulse excitation in the excited state $| e_{\rm A}, m = +1 \rangle$. Then atom $A$ will emit
a $\sigma^+$ polarized photon and drop to the ground state $|g_{\rm A} \rangle$. Clearly this
$\sigma^+$ photon can excite the transition $|g_{\rm B} \rangle \rightarrow | e_{\rm B}, m= +1 \rangle$.
We use a fixed coordinate system as shown in Fig.~\ref{Fig:Atoms}. The quantization axis is
taken to be perpendicular to the PS. We introduce the circular polarization
vector $\boldsymbol{\epsilon}_\pm = (\mathbf{e}_x \pm \ri \mathbf{e}_y)/\sqrt{2}$ where 
$\mathbf{e}_x$ and $\mathbf{e}_y$  are the unit vectors in $x$ and $y$ direction, respectively.
Now, we ask the question --- can the $\sigma^+$ photon emitted by atom $A$ excite the transition
$|g_{\rm B} \rangle \rightarrow | e_{\rm B}, m= -1 \rangle$ in atom $B$? Clearly this
requires a $\sigma^-$ photon defined with respect to the fixed coordinate system. This
would be possible if the propagation of the electromagnetic field is such that the $\sigma^+$
photon gets partially converted into a $\sigma^-$ photon when it reaches atom $B$. This
conversion would be due to the spin-orbit interaction inherent in Maxwell equations. Let
us denote by $l$ the orbital angular momentum then if $\sigma^+$ is converted into $\sigma^-$
such a conversion would happen by producing two units of $l$, i.e.\ 
$|\sigma^+, l = 0\rangle \rightarrow |\sigma^-, l = 2 \rangle$. Thus the excitation of atom $B$
to the state $| e_{\rm B}, m = -1 \rangle$ is the SHEL in FRET. It should be borne in mind that
conversion of $\sigma^+$ to $\sigma^-$ would happen with a certain probability and hence the probability
of exciting atom $B$ to the state $| e_{\rm B}, m = -1 \rangle$ is nonzero. The SHEL in FRET
can be monitored by studying the populations of the excited states by applying another field and
by studying far off resonant flourescence.

We next present our method of calculation. The Hamiltonian in the interaction picture can be written as
in the interaction picture by the Hamiltonian
\begin{equation}
  H_1(t) = - \sum_{\alpha,j} \mathbf{p}_{\alpha,j}\cdot\mathbf{E}(\mathbf{r_\alpha},t) |e_{\alpha,j} \rangle \langle g_\alpha | \re^{\ri \omega_0 t} + {\rm h.c.}
\end{equation}
Here the sum is over both the atoms ($\alpha = A,B$) and the two excited states of each atom. Note
that $\langle e, m = \pm 1| \mathbf{p} |g\rangle = p \boldsymbol{\epsilon}_\pm^*$ 
where $p$ is the amplitude of the dipole matrix element. Further $\mathbf{E}$ is the Hermitian
electromagnetic field operator. We have four possibilities of energy transfer: 
two with $\Delta m = 0$ ($|e_{A},m=+1\rangle \rightarrow |e_{B},m=+1\rangle$, $|e_{A},m=-1\rangle \rightarrow |e_{B},m=-1\rangle$) 
and two with $\Delta m = \pm 2$ ($|e_{A},m=-1\rangle \rightarrow |e_{B},m=+1\rangle$, $|e_{A},m=+1\rangle \rightarrow |e_{B},m=-1\rangle$).

If we assume that atom $A$ is prepared in an excited state 
as depicted in Fig.~\ref{Fig:Atoms} (a) then the wave function of the system at time $t$ is 
up to second order in $H_1$ given by
\begin{equation}
  | \psi(t) \rangle = -\frac{1}{\hbar^2} \int_{-\infty}^t \!\!\!\!\rd t_1 \int_{-\infty}^t \!\!\!\!\rd t_2\, H_1(t_1) H_1(t_2)|\psi_0\rangle 
\end{equation}
where $|\psi_0 \rangle = |e_A, m= +1, g_B,\{0\} \rangle$; $\{0\}$ denotes the vacuum state of the electromagnetic field in the presence of the PS.

We are now interested in the probability of exciting atom $B$ at a later time. Therefore
we need to calculate the transition amplitude  $C = \langle g_A, e_B, \{0\} | \psi(t) \rangle$.
It can be shown that~\cite{AgarwalBook}
\begin{equation}
  \frac{\partial C}{\partial t} = \frac{\ri \omega_0^2}{\hbar c^2} \mathbf{p}_B \cdot \mathds{G}(\mathbf{r}_B,\mathbf{r}_A,\omega_0) \cdot \mathbf{p}_A^*
\label{Eq:TransitionAmplitudePartial}
\end{equation}
where the asterisk symbolizes the complex conjugation. 
Here, we have introduced the dyadic Greens function $\mathds{G}$ of
the environment. Note that the full calculation leading to Eq.~(\ref{Eq:TransitionAmplitudePartial})
includes both resonant and nonresonant quantum paths 
$|e_{\rm A}, g_{\rm B} \rangle \rightarrow |g_{\rm A}, g_{\rm B} \rangle \rightarrow |g_{\rm A}, e_{\rm B} \rangle$;
$|e_{\rm A}, g_{\rm B} \rangle \rightarrow |e_{\rm A}, e_{\rm B} \rangle \rightarrow |g_{\rm A}, e_{\rm B} \rangle$.
Eq.~(\ref{Eq:TransitionAmplitudePartial}) gives the probability which can be converted in the
usual manner to the transition rate by summing over the final states. We normalize the transition
rate by the Einstein A coefficient $\Gamma = \frac{4 \omega_0^3 |p|^2}{3 \hbar c^3}$ for
the level $|e_{\rm A}, m = +1\rangle$. We call this ratio as $D_{\rm n}$ [for FRET to $| e_{\rm B}, m = +1 \rangle$]
and $D_{\rm sh}$  [for FRET to $| e_{\rm B}, m = -1 \rangle$]
\begin{align}                                                                   
  D_{\rm sh} &= \frac{9}{32} \biggl| \frac{\bigl( \mathds{G}_{xx} - \mathds{G}_{yy} + \ri \bigl[ \mathds{G}_{xy} + \mathds{G}_{yx} \bigr] \bigr)}{\omega_0/c} \biggr|^2, \label{Eq:Dsh} \\
  D_{\rm n}  &= \frac{9}{32} \biggl| \frac{\bigl( \mathds{G}_{xx} + \mathds{G}_{yy} \bigr)}{\omega_0/c} \biggr|^2.\label{Eq:Dn}
\end{align}
It is important to note that we have finite probabilities for both kinds
of excitation transitions which refer to a zero spin-angular momentum 
coupling ($\Delta m = 0$) or a non-zero spin-angular momentum 
coupling ($\Delta m = \pm 2$).

Let us first consider the process of FRET in free space. Then we can determine
the transition probabilities by means of the free Green's function $\mathds{G}^{(v)}$ which is
given by
\begin{equation}
  \mathds{G}_{ij}^{(v)}(\mathbf{r,r'},\omega) = \biggl( \delta_{ij} + \frac{c^2}{\omega^2} \frac{\partial^2}{\partial r_i \partial r_j} \biggr) \frac{\re^{\ri \frac{\omega}{c} |\mathbf{r - r'}|}}{|\mathbf{r-r'}|}.
\end{equation}
When assuming that both atoms are in a plane perpendicular to the $z$ axis defined by the
quantization axis and let $d$ be the interatomic distance, then we obtain
\begin{align}                                                                   
   D_{\rm sh}^{(v)} &= \frac{9}{32} \biggl|\frac{\re^{\ri k_0 d}}{k_0 d} b \re^{+ 2 \ri \varphi} \biggr|^2, \label{Eq:C2} \\ 
   D_{\rm n}^{(v)} &= \frac{9}{32}  \biggl|\frac{\re^{\ri k_0 d}}{k_0 d} (2 a + b)\biggr|^2, \label{Eq:C0}
\end{align}
where we have introduced $a = (k_0^2 d^2 + \ri k_0 d - 1)/k_0^2 d^2$, and $b = (3 - 3 \ri k_0 d - k_0^2 d^2)/k_0^2 d^2$;
$k_0 = \omega_0 / c$. The angle $\varphi$ appearing in the first expression is the angle between the
line connecting the positions of the atoms $A$ and $B$ with the $x$-axis. Eq.~(\ref{Eq:C2}) shows that during the excitation transition
$D_{\rm sh}^{(v)}$ the orbital angular momentum has changed by two units which is a manifestation of 
the spin-orbit interaction.

\begin{figure}[Hhbt]
  \epsfig{file = 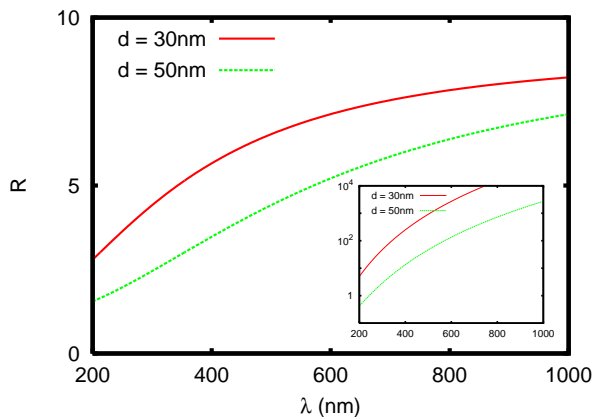, width = 0.45\textwidth}
  \caption{\label{Fig:Vacuum} $R$ from Eq.~(\ref{Eq:VacuumEnhancement}) as function of the wavelength. The inset 
           shows a semi-logarithmic plot of $D_{\rm sh}^{(v)}$ from Eq.~(\ref{Eq:C2}) for the same wavelength range.}
\end{figure}

To see which of both excitation transitions is more likely, we plot in Fig.~\ref{Fig:Vacuum} the 
relative transition probability 
\begin{equation}
  R = \frac{D_{\rm sh}^{(v)}}{ D_{\rm n}^{(v)}} = \frac{|b|^2}{|2 a + b|^2}
\label{Eq:VacuumEnhancement}
\end{equation}
for free space. It can be seen that for the considered wavelength
region and the interatom distances $d = 30\,{\rm nm}$ and $d = 50\,{\rm nm}$ the 
SHEL is prominent, i.e.\ energy transfer with change in spin-angular momentum 
is generally speaking more likely than the one with no change in spin-angular momentum. 
Note also that for small distances $k_0 d \ll 1$ we have $a \approx -1/(k_0 d)^2$
and $b \approx 3/(k_0 d)^2$ so that $R \approx 9$; for large distances $k_0 d \gg 1$
we find $R \approx 1$.

Now, we want to show that the SHEL in free space can be enhanced up to several orders
of magnitude by the interaction of the two atoms $A$ and $B$ with the surface plasmons of a given
sample. To this end, we first need to determine $D_{\rm sh}$ and $D_{\rm n}$ for the case where the atoms
are placed in a given distance $z$ above a PS (see Fig.~\ref{Fig:Atoms}). 
This can be done by using a spacer material
on the sample, for instance. Then the Green's function is given by the sum 
$\mathds{G} = \mathds{G}^{(v)} + \mathds{G}^{(sc)}$ of the vacuum Green's function $\mathds{G}^{(v)}$ 
and the scattered Greens function $\mathds{G}^{(sc)}$ which takes the interaction 
with the surface into account. Analytical expressions for $\mathds{G}^{(sc)}$ for a sample
with a plane surface can be found in Ref.~\cite{NovotnyBook}. When inserting the full Greens function
into Eqs.~(\ref{Eq:Dsh}) and (\ref{Eq:Dn}) we obtain 
\begin{align}
   D_{\rm sh} &= \frac{9}{32} \biggl| \re^{+2 \ri \varphi} \int_0^\infty \!\!\frac{\rd \kappa}{2 \pi} \, \frac{\ri \kappa}{2 k_0 \gamma_0}   J_2(\kappa d) \nonumber \\ 
    &\qquad \times \biggl[ \bigl(1 + r_\rs \re^{2 \ri \gamma_0 z}\bigr) 
     -  \frac{\gamma_0^2}{k_0^2} \bigl(1 + r_\rp \re^{2 \ri \gamma_0 z}\bigr) \biggr]\biggr|^2, \label{Eq:SHELPS} \\
   D_{\rm n} &=  \frac{9}{32} \biggl| \int_0^\infty \!\!\frac{\rd \kappa}{2 \pi} \,  \frac{\ri \kappa}{2 k_0 \gamma_0}  J_0(\kappa d) \nonumber\\
    & \qquad\times \biggl[ \bigl(1 + r_\rs \re^{2 \ri \gamma_0 z}\bigr) 
    +  \frac{\gamma_0^2}{k_0^2} \bigl(1 + r_\rp \re^{2 \ri \gamma_0 z}\bigr) \biggr]\biggr|^2,\label{Eq:NPS}
\end{align}
where we have introduced $\gamma_0 = \sqrt{k_0^2 - \kappa^2}$ the wavevector in $z$ direction, 
$\boldsymbol{\kappa} = (k_x,k_y)^t$ the wavevector parallel to the surface; $J_0$ and $J_2$
are the cylindrical Bessel functions. $r_{\rm s}$ and $r_{\rm p}$ are the amplitude reflection coefficients
of the PS for s and p polarization. The poles of $r_{\rm p}$ determine the surface modes of the PS. 
Further, Eqs.~(\ref{Eq:SHELPS}) and (\ref{Eq:NPS}) are far from paraxial, since all possible $\kappa$ are taken into account: 
the propagating waves ($\kappa \in [0,k_0]$) as well as the evanescent waves ($\kappa \in [k_0, \infty [$). 
This time the coupling to the angular momentum manifests itself not only through the phase factor 
$\re^{\mp 2 \ri \varphi}$ but also through the different Bessel functions.

\begin{figure}[Hhbt]
  \epsfig{file = 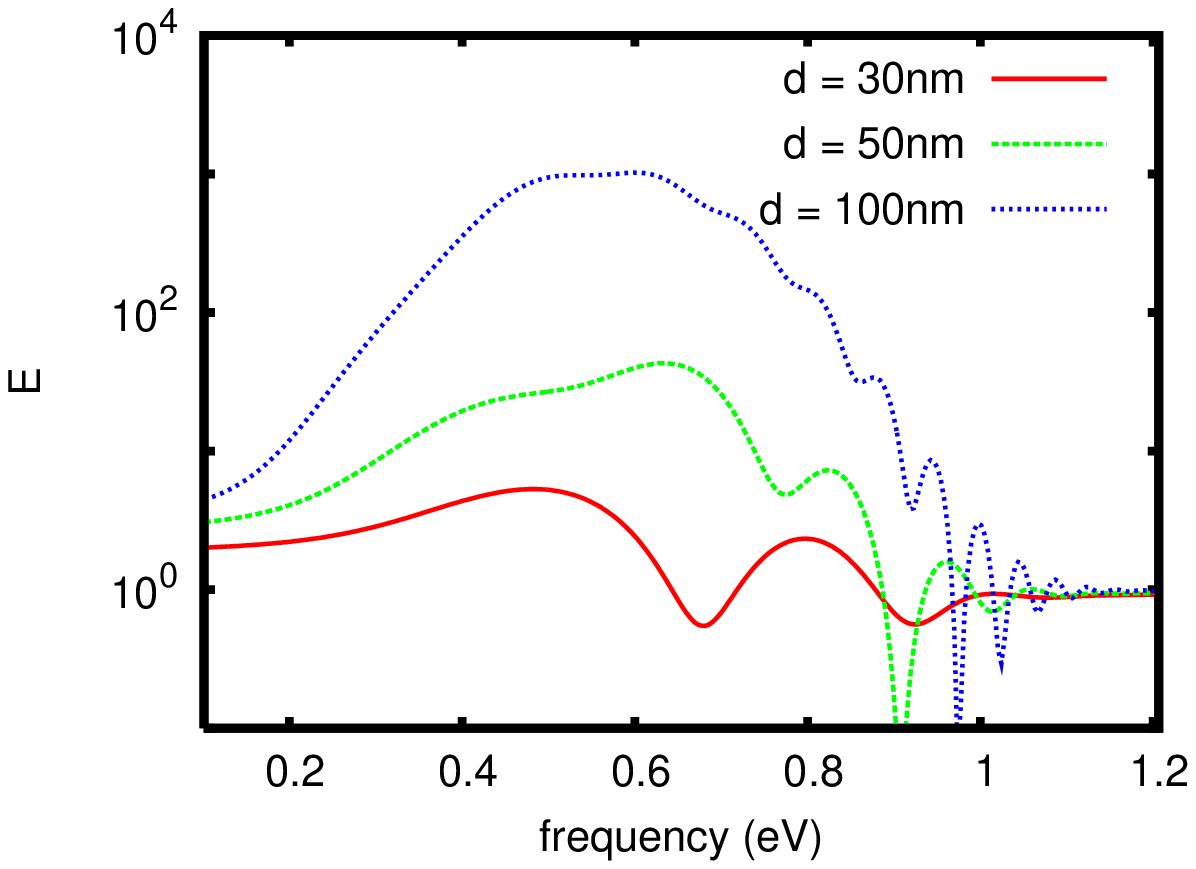, width = 0.45\textwidth}
  \epsfig{file = 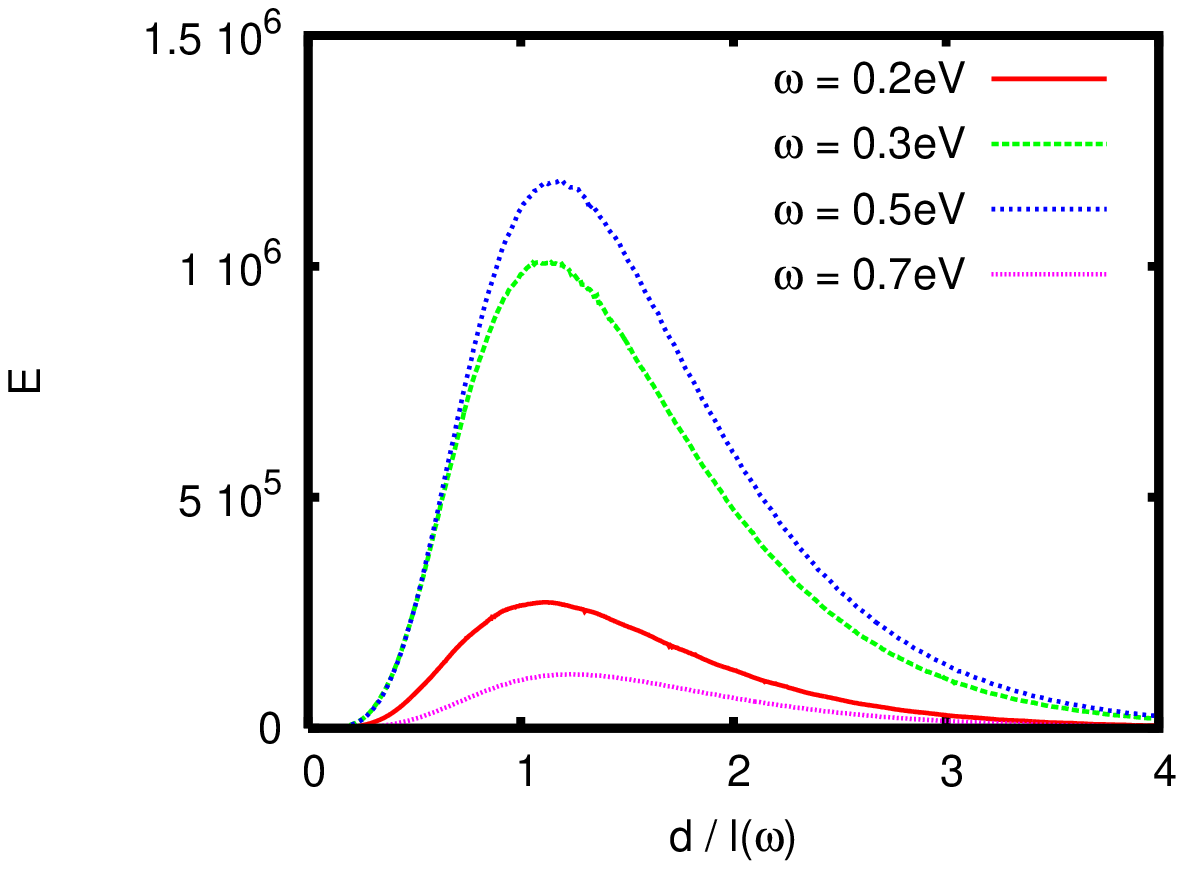, width = 0.45\textwidth}
  \caption{(a) The giant SHEL on suspended graphene with $E_{\rm F} = 1\,{\rm eV}$ as function of wavelength. 
           (b)  $E$ as function of distance. The distance is normalized by the 
                corresponding propagation length of the surface mode $l = 1/\Im(\kappa_{\rm sp})$ which is
                $l(0.2\,{\rm eV}) = 2138\,{\rm nm}, l(0.3\,{\rm eV}) = 1392\,{\rm nm},  l(0.5\,{\rm eV}) = 772\,{\rm nm}$,
                and $  l(0.7\,{\rm eV}) = 486\,{\rm nm}$.}
\label{Fig:Graphene}
\end{figure}

Finally, we introduce the plasmonic enhancement factor
\begin{equation}
  E \equiv \frac{D_{\rm sh}}{D_{\rm sh}^{(v)}}
\end{equation}
which allows us to quantify the enhancement of the SHEL by means of the interaction
with the surface plasmons of the sample. As a first example let us consider a suspended 
sheet of graphene. The reflection coefficient for the p-polarized modes is in this case 
given by~\cite{KoppensEtal2011}
\begin{equation}
   r_\rp = \frac{\frac{4 \pi \sigma \gamma_0}{\omega}}{2 + \frac{4 \pi \sigma \gamma_0}{\omega}},
\label{Eq:ReflGraphene}
\end{equation}
Using this quantity with the analytical expressions 
for the conductivity $\sigma(\omega)$ of graphene~\cite{KoppensEtal2011}
we obtain the plasmonic enhancement $E$ plotted in Fig.~\ref{Fig:Graphene}. In Fig.~\ref{Fig:Graphene}(a)
one can observe the frequency dependence showing a broadband effect of plasmonic enhancement due to
the interactions of the plasmons in the graphene sheet which exist for frequencies smaller than $1\,{\rm eV}$
for the chosen doping level. In Fig.~\ref{Fig:Graphene}(b) we show the distance dependence
of the plasmonic enhancement choosing different frequencies. The curves are normalized to the corresponding
propagation length of the surface plasmons $l = 1/\Im(\kappa_{\rm sp})$ which can be derived from the
poles of the reflection coefficient in Eq.~(\ref{Eq:ReflGraphene}). Obviously, the enhancement dies off 
for interatom distances $d \gg l$ which shows that the plasmonic enhancement is due to a excitation 
transition mediated by the surface plasmons. Note, that for certain distances the plasmonic enhancement leads
to a transition probability which is six orders of magnitude larger than for vacuum. 
That the SHEL is so large can be explained by the fact that due to the 
interaction with the surface plasmons of the PS the emission of the photon by atom 
$A$ is enhanced as well as the absorption by atom $B$. Thus local field enhancement factors contribute 
twice making th SHEL {\itshape giant}.

\begin{figure}[Hhbt]
  \epsfig{file = 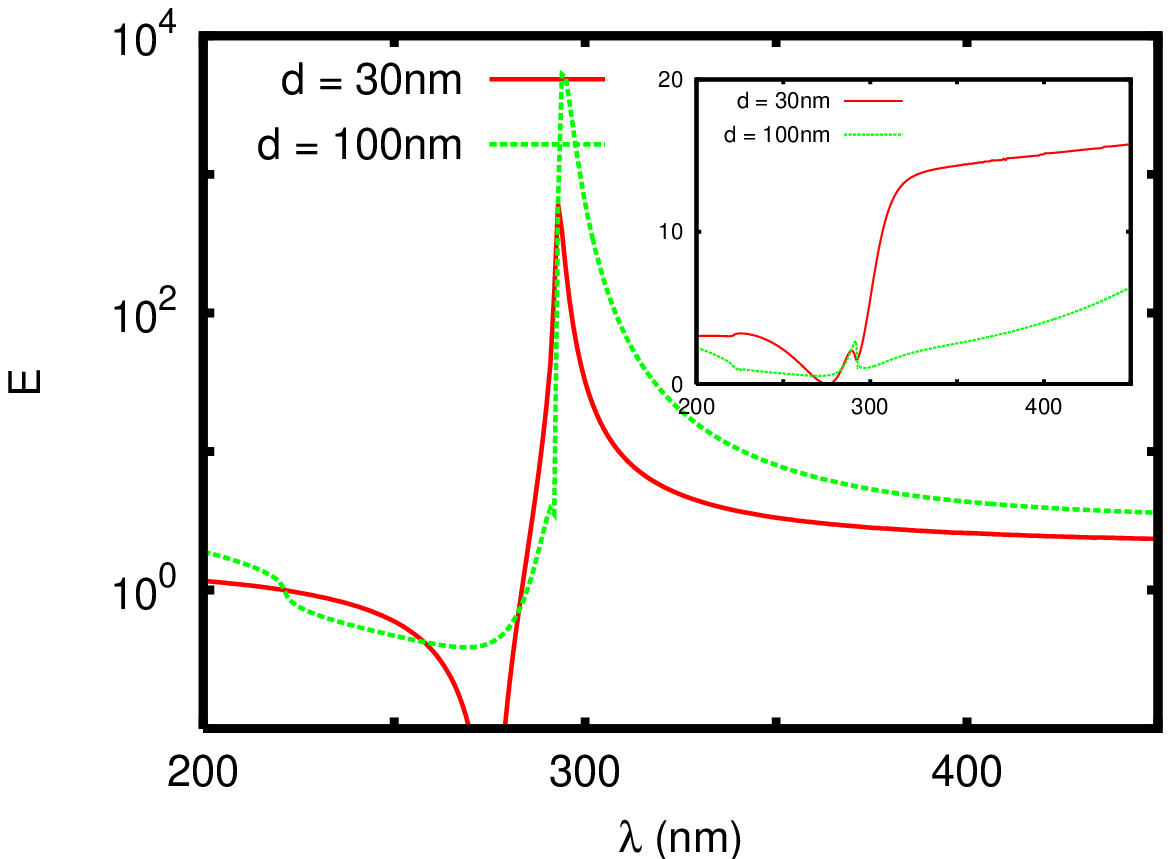, width = 0.45\textwidth}
  \epsfig{file = 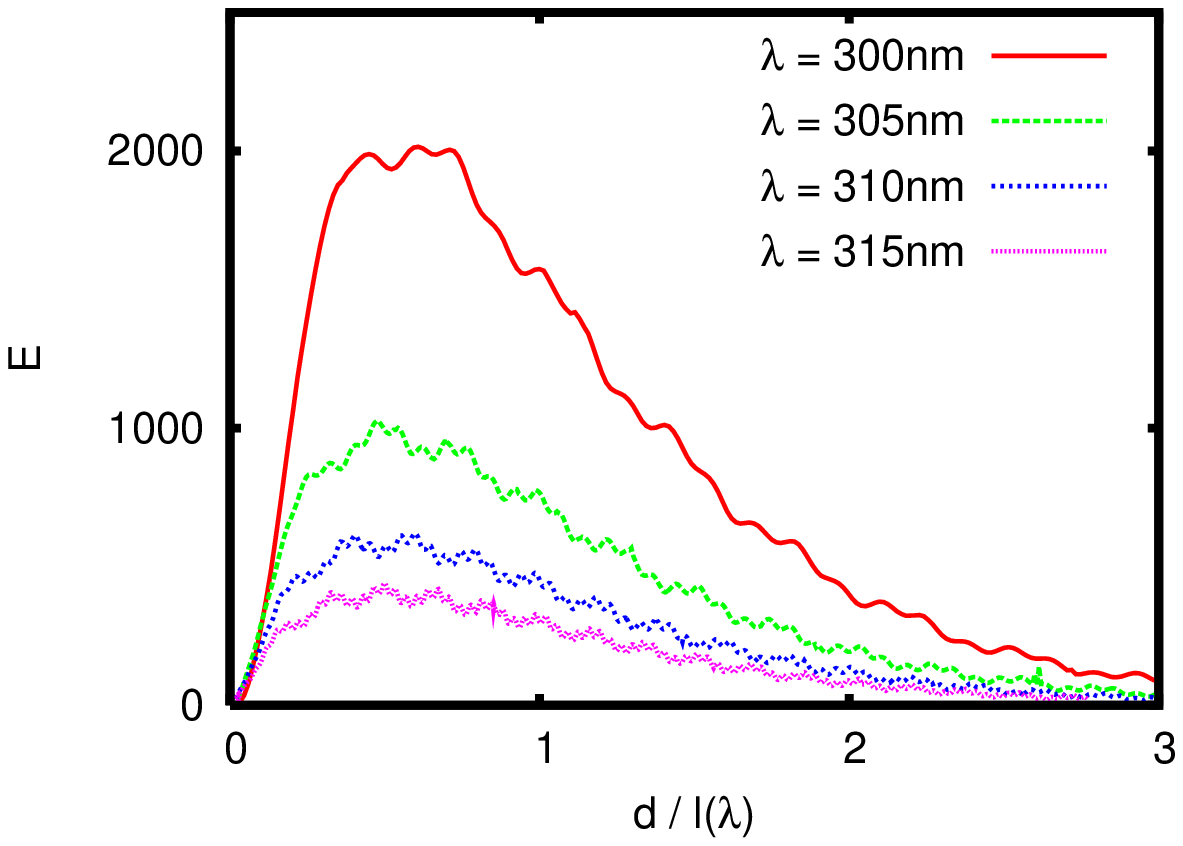, width = 0.45\textwidth}
  \caption{(a) $E$ for semi-infinite silver as function of wavelength. The inset 
               shows $D_{sh}/D_n$ for the same distances and wavelength range.
           (b) $E$ for semi-infinite silver for different wavelengths close to 
               the surface mode resonance as function of distance d. The distance is normalized the corresponding propagation 
               length of the surface mode $l = 1/ \Im(\kappa_{\rm sp})$ which is $l(300,{\rm nm}) = 692\,{\rm nm}, l(305,{\rm nm}) = 1458\,{\rm nm}$, $l(310,{\rm nm}) = 2436\,{\rm nm}$, and $l(315,{\rm nm}) = 3610\,{\rm nm}$.}
\label{Fig:Silver}
\end{figure}

Next we consider the role of surface plasmons on a semi-infinite metal surface.
In this case the reflection coefficient is given by the usual Fresnel coefficient
\begin{equation}
   r_{\rm p} = \frac{\gamma_0 \epsilon - \gamma}{\gamma_0 \epsilon + \gamma}
\label{Eq:ReflSilver}
\end{equation}
where $\epsilon$ is the permittivity of the sample and $\gamma = \sqrt{k_0^2 \epsilon - \kappa^2}$. 
Here, we use a Drude model~\cite{Soennichsen} which has been fitted to the
Johnson-Christy data~\cite{ChristyJohnson}. The resulting plasmonic enhancement $E$ is plotted in Fig.~\ref{Fig:Silver}.
The wavelength dependence in Fig.~\ref{Fig:Silver}(a) shows a large enhancement for wavelenght of $\lambda_{\rm sp} = 291,7\,{\rm nm}$.
This is exactly the surface plasmon resonance frequency of silver which follows from $\Re[\epsilon(\omega_{\rm sp})] = -1$ 
and $\Im[\epsilon(\omega_{\rm sp})] \ll 1$. Here, the observed plasmonic enhancement results in a probability
$D_{\rm sh}$ being four orders of magnitude larger than for vacuum. In addition, this is  
a narrow-band effect compared with graphene which is due to the fact that for the chosen distance $z$ we are 
in the electrostatic regime of the surface mode.  We have also determined
the pure pole contribution to $D_{\rm sh}$ which describes the narrow peak in Fig. 4(a).
Note also the Fano-like line shapes for SHEL which are due to the interference of the contributions 
from the pole of $r_{\rm p}$ and the non-pole contribution in Eq.~(\ref{Eq:SHELPS}).
In Fig.~\ref{Fig:Silver}(b) we plot the plasmonic enhancement 
factor as a function of the interatom distance $d$. Again, each curve is normalized to the propagation length of 
the surface plasmons in silver $l = 1/\Im(\kappa_{\rm sp})$ which is now obtained from the poles of 
Eq.~(\ref{Eq:ReflSilver}). Once more, one can observe that the plasmonic enhancement dies off for $d \gg l$.

\begin{figure}[Hhbt]
  \epsfig{file = 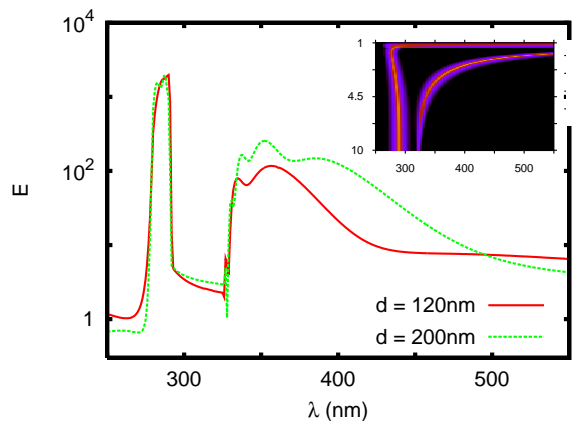, width = 0.45\textwidth}
  \caption{$E$ for a $15\,{\rm nm}$ silver film on silica as function of wavelength. 
           The inset shows a 3-d plot of $\log\bigl( \Im[r_{\rm p} (\omega,\kappa) ] \bigr)$ from Eq.~(\ref{Eq:ReflSilver}) 
           for the same wavelength range and lateral wavevectors $\kappa/k_0 \in [1, 10]$.}
\label{Fig:SilverFilm}
\end{figure}

As in studies of spontaneous emission~\cite{RussellEtAl2012,LeeEtAl2011,NealEtAl2011}, the possibility of 
tayloring the properties of the surface modes will affect the plasmonic enhancement of the SHEL as well. 
This can for example be done by sputtering thin metallic films on a dielectric substrate. Then the surface 
mode branch will split into two surface mode branches~\cite{Economou1969}. This splitting should also be 
obsverable in the Spin-Hall effect. In Fig.~\ref{Fig:SilverFilm} we show the plasmonic enhancement factor
for two atoms above a substrate made of a $15\,{\rm nm}$ silver film on a silica substrate 
($\epsilon_{\rm substrate} = 2.1025$). The splitting of the surface plasmons can be nicely seen.
Furthermore, it can be observed that the excitation transfer mediated by the long range plasmon~\cite{Sarid1981} 
($\lambda <300\,{\rm nm}$) leads to a much larger enhancement than that by the
short range plasmon ($\lambda > 320\,{\rm nm}$) which is similar to the enhanced emission rate
of a molecule close to a corrugated metal film~\cite{LeungEtAl1989}.

In conclusion we have presented a new way to realize the spin Hall effect of light which is inherent in the near field of an oscillating dipole. We showed how this kind of SHEL can be monitored by observing the energy transfer to an orthogonally polarized atomic state. The change in polarization is accompanied by the generation of two units of orbital angular momentum. The SHEL is largely enhanced on plasmonic platforms due to resonant surface plasmons. Other systems like nano fibers or nano antennas could be equally useful in producing large SHEL.  

The idea was initially presented by GSA at the TIFR school on Plasmonics 2012 in Hyderabad and GSA thanks the Director TIFR for hospitality.

\end{document}